\documentstyle[prl,floats,aps]{revtex}
\begin{document}
\draft
\twocolumn
\begin{title}
{ Nonadiabatic superconductivity in MgB$_{2}$ and cuprates}
\end{title} 
\author{ A.S. Alexandrov}
\address{Department of Physics, Loughborough University,
Loughborough LE11 3TU, UK}

\maketitle
\begin{abstract}
The Fermi energy, $E_{F},$ of newly discovered high-$T_{c}$ superconductor
MgB$_{2}$ and cuprates is estimated from the measured London penetration
depth using a parameter-free expression. $E_{F}$ of MgB$_{2}$ and
more than 30 lanthanum, yttrium and mercury-based cuprates appears to be
about or below 100 meV, depending on doping. There is every evidence
that
the remarkable
low value of $E_{F}$ and the strong coupling of carriers with high-frequency
phonons is the cause of high $T_{c}$ in all newly discovered
superconductors. Taking into account that the carriers mainly reside on
oxygen in cuprates, on boron in magnesium diborade, and on carbon in doped
fullerenes, these superconductors form what is essentially $nonadiabatic$
'metallic' oxygen, boron and carbon, respectively.  The boron
isotope effect on the carrier mass in magnesium-diboride similar to the
oxygen isotope effect on the supercarrier mass observed by Guo-meng
Zhao $et$ $al.$ in cuprates is predicted.
\end{abstract}

\narrowtext

\vspace{0.5cm}

One of the common features of all newly discovered high-temperature
superconductors (T$_{c}\geq 30K$) is a strong coupling of carriers with
optical phonons, as follows from the experimental values of their static and
high-frequency dielectric constants \cite{alebra}, other optical probes \cite
{mic}, isotope effect \cite{zao,isoMg}, photoemission \cite{she}, tunnelling 
\cite{pon}, and first-principles band structure calculations. Hence, it is
natural to expect that the Migdal-Eliashberg \cite{mig,eli} BCS-like
approach to coupled system of electrons and phonons would account for their
normal state and superconducting properties. The approach is based on
``Migdal's theorem'' \cite{mig}, which showed that the contribution of the
diagrams with 'crossing' phonon lines (so called 'vertex' corrections) is
small if the parameter $\lambda \omega /E_{F}$ is small ($\hbar =k_{B}=c=1$%
). Neglecting the vertex corrections, Migdal calculated the renormalized
electron mass as $m^{\ast }=m(1+\lambda )$ (near the Fermi level) \cite{mig}
and Eliashberg later extended the Migdal theory to describe the BCS
superconducting state at intermediate values of the electron-phonon (BCS)
coupling constant $\lambda $ \cite{eli} by breaking the gauge symmetry. The
same theory, applied to phonons, yields the renormalized phonon frequency $%
\tilde{\omega}=\omega (1-2\lambda )^{1/2}$ \cite{mig} with an instability at 
$\lambda =0.5$. Because of this instability, both Migdal \cite{mig} and
Eliashberg \cite{eli} restricted the applicability of their approach to $%
\lambda \leq 1$. It was then shown that if the adiabatic Born-Oppenheimer
approach is properly applied to a metal, there is only negligible
renormalization of the phonon frequencies of the order of the adiabatic
ratio, $\omega /E_{F}\ll 1$ at $any$ value of $\lambda $. The conclusion was
that the standard electron-phonon interaction could be applied to electrons
for any value of $\lambda $, but it should not be applied to renormalize the
characteristics of the phonons \cite{gei}. As a result, many authors used
the Migdal-Eliashberg theory with $\lambda $ much larger than $1$ (see, for
example, Ref.\cite{sca}).

However, starting from the infinite coupling limit, $\lambda =\infty ,$ and
applying the inverse ``$1/\lambda $'' expansion technique \cite{fir} it was
shown \cite{ale} that a many-electron system collapses into
small (nonadiabatic) polaron regime at $\lambda \geq 1$ almost independently
of the adiabatic ratio. This regime is beyond the Migdal-Eliashberg theory.
It cannot be reached by summation of the standard Feynman-Dyson perturbation
diagrams with the translation-invariant Green's functions even including $%
all $ vertex corrections, because of the broken translational symmetry, as
first discussed by Landau \cite{lan} for a single electron and by  us 
\cite{ale} for the many-electron system. In recent years quite a few
numerical and analytical studies have confirmed this conclusion \cite
{kab,kab2,bis,feh,mar,alemot,tak,dev,feh2,tak2,rom,lam,zey,wag,bon2,alekor,aub2,ale2}
(and references therein). Of course, if the coupling is not very strong ($%
\lambda <1$) $and$ the adiabatic ratio is sufficiently small, $\omega
/E_{F}\ll 1$, one can apply the standard Feynman-Dyson perturbation
technique including the vertex corrections to the vertex function \cite
{all,pie}. If one or both of these two conditions is not satisfied then the
nonadiabatic polaron theory of superconductivity \cite{aleF} is more
appropriate. 

Here I show that the adiabatic ratio in all novel
superconductors is of the order or even larger than unity for most essential
optical phonons.

Consider recent high-temperature superconductor MgB$_{2}$ \cite{dis}. The
crystal structure of magnesium-diboride contains planes of magnesium and
graphite-like planes of boron atoms. First-principles band structure
calculations \cite{boy,pic} show that magnesium donates its outer electrons
to boron layers like copper donates its electrons to oxygen in cuprates, and
alkali metals to carbon in doped C60. The band structure is
quasi-two-dimensional with two generate hole pockets \cite{pic} and the $%
bare $ (LDA) Fermi energy about $0.55$ eV. Applying the parabolic
approximation for the band dispersion one obtains the renormalized Fermi
energy as 
\begin{equation}
E_{F}={\frac{\pi n_{i}d}{{m_{i}^{\ast }}}},
\end{equation}
where $d$ is the interplane distance, and $n_{i},m_{i}^{\ast }$ are the
density of holes and their effective mass in each of two bands ($i=1,2$),
renormalized due to the electron-phonon (and electron-electron) interaction.
One can express the renormalized band-structure parameters through the
in-plane London penetration depth at $T=0$, measured experimentally: 
\begin{equation}
\lambda _{H}^{-2}=4\pi e^{2}\sum_{i}\frac{n_{i}}{m_{i}^{\ast }}.
\end{equation}
As a result, one obtains the {\em parameter-free }expression for the
``true'' Fermi energy as 
\begin{equation}
E_{F}={\frac{d}{{4ge^{2}\lambda _{H}^{2}}}},
\end{equation}
where $g$ is the degeneracy of the spectrum ($g=2$ in MgB$_{2}$). The same
expression applies in cuprates because of the two-dimensional character of
their band structure. However, the degeneracy $g$ in cuprates may depend on
doping. In underdoped cuprates one expects 4 hole pockets inside the
Brillouin zone (BZ) due to the Mott-Hubbard gap. If the hole band minima are
shifted with doping to BZ boundaries, all their wave vectors would belong to
the stars with two or more prongs. The groups of the wave vector for this
stars have only 1D representations. It means that the spectrum will be
degenerate with respect to the number of prongs which has the star, i.e $%
g\geq 2$. The only exception is the minimum at ${\bf k}=(\pi ,\pi )$ with
one prong and $g=1$. Hence, in the cuprates the degeneracy is $1\leq g\leq 4$%
.

Generally, the ratios $n/m$ in Eq.(1) and Eq.(2) are not necessary the
same. The `superfluid' density in Eq.(2) might be different from the total
density of delocalized carriers in Eq.(1). However, in a translationally
invariant system they must be the same \cite{leg}. This is true even in the
extreme case of a pure two-dimensional 
superfluid, where quantum fluctuations are important \cite{lok}. One can,
however, obtain a reduced value of the zero temperature superfluid density
in the dirty limit, $l\ll \xi (0)$, where $\xi (0)$ is the zero-temperature
coherence length. The latter was measured directly in cuprates as the size
of the vortex core. It is about 10 $\AA $ or even less. On the contrary, the
mean free path was found surprisingly large at low temperatures, $l\sim $
100-1000 $\AA $. Hence, I believe that all novel superconductors,
including MgB$_2$ are in the
clean limit, $l\gg \xi (0)$, so that the parameter-free expression for $%
E_{F} $, Eq.(3), is perfectly applicable.

Parameter-free estimate of the Fermi energy, $E_{F},$ obtained by using
Eq.(3), is presented in the Table. It becomes clear that the Fermi energy in
magnesium diboride and in more than 30 cuprates is about 100 meV or
even less, in particular, if the degeneracy $g\geq 2$ is
taken into account. That should be compared with the characteristic phonon
frequency, which can be estimated as the plasma frequency of boron or oxygen
ions, 
\begin{equation}
\omega =(4\pi Z^{2}e^{2}N/M)^{1/2}.
\end{equation}
With $Z=1$, $N=2/V_{cell}$, M=10 a.u. one obtains $\omega \simeq $ 69 meV for MgB$_{2}$, and $\omega $=84meV with $Z=2$, $%
N=6/V_{cell}$, $M$=16 a.u. for YBa$_{2}$Cu$_{3}$O$_{6}$. Here $V_{cell}$ is
the volume of the (chemical) unit cell. The estimate agrees well with the
measured phonon spectra \cite{sat,tim}. As established experimentally in
cuprates \cite{tim}, the high-frequency
phonons are strongly coupled with carriers. The parameter-free expression,
Eq.(3), does not apply to doped fullerenes with their three-dimensional
energy structure. However, it is well established that they are also in the
nonadiabatic regime \cite{alekabF}.

The low Fermi energy (Table), $E_{F}\leq \omega $ is a serious problem
within the Migdal-Eliashberg approach. In the framework of this BCS-like
approach (largely independent of the nature of coupling) the critical
temperature is fairly well approximated by McMillan's formula (see in e.g.
Ref. \cite{sca}), 
\begin{equation}
T_{c}={\frac{\omega }{{1.45}}}\exp \left[ -{\frac{1.04(1+\lambda )}{{\lambda
-\mu ^{\ast }(1+0.62\lambda )}}}\right] ,
\end{equation}
which works well for simple metals and their alloys. There are no general
restrictions on the BCS value of T$_{c}$ if the dielectric function
formalism is properly applied \cite{max}. Allen and Dynes \cite{all2} found
that in the strong-coupling limit $\lambda \gg 1$ the critical temperature
might be as high as $T_{c}\simeq \omega \lambda ^{1/2}/2\pi $. Nevertheless,
applying this kind of theory to novel superconductors is problematic. Since
the Fermi energy is small and phonon frequencies are high, the Coulomb
pseudopotential $\mu ^{\ast }$ is of the order of the bare Coulomb
repulsion, $\mu ^{\ast }\simeq \mu \simeq 1$ because the
Tolmachev-Morel-Anderson logarithm is ineffective. Hence, to get
experimental T$_{c}$ with Eq.(5), one has to have a strong coupling, $%
\lambda \gg 1$. However, one cannot increase $\lambda $ without accounting
for the polaron collapse of the band. As discussed above, this happens at $%
\lambda \simeq 1$ for uncorrelated polarons, and even for a smaller value of
the bare electron-phonon coupling in strongly correlated models \cite{feh}.
Of course, one can argue \cite{max2}, that a renormalized value of the
coupling $\tilde{\lambda}\sim \lambda /(1-2\lambda )$ appears in Eq.(5),
rather than a bare $\lambda $ because of the familiar Migdal's softening of
the phonon spectrum. That leaves some room for high T$_{c}$ in the region of
the applicability of the Eliashberg theory (i.e. $\lambda \leq 0.5$).
However, even in this region the non-crossing diagrams cannot be treated as
vertex $corrections$ because $\omega /E_{F}\geq 1$, since they are
comparable to the standard terms.

To conclude, I have shown that MgB$_{2}$ and cuprates are in the
nonadiabatic regime, where the Migdal-Eliashberg theory is inappropriate.
The interaction with optical high-frequency phonons should be treated within
the multi-polaron theory \cite{aleF} based on the canonical Lang-Firsov
transformation. The renormalized Fermi energy is  one order of
magnitude lower than the bare (LDA) Fermi energy in MgB$_{2}$, which
corresponds to ca 2 phonons dressing a hole \cite{alemot}. The 'bare' phonon dressing  might be
even stronger because
the electron correlations  undress  polarons,
as argued by Hirsh \cite{hir}.
Nevertheless, the ground state of MgB$_{2}$ might yet be a (polaronic) BCS/
Fermi liquid \cite{ale} if the interaction with these phonons is not so
strong that the bipolaron (real-space) pairs would form. Then, as in the
case of doped fullerenes \cite{alekabF}, the interaction with low-frequency
phonons ($\omega \ll E_{F}$) could be accounted for within the framework of
the Migdal-Eliashberg approach. While the Fermi surface topology is not
changed in the canonical transformation, there is a qualitative difference
between the ordinary Fermi liquid and the nonadiabatic polarons. In
particular, the renormalized (effective) mass of electrons is independent of
the ion mass $M$ in ordinary metals (where the Migdal approximation is
believed to be valid), because $\lambda $ does not depend on the isotope
mass. However, the polaron effective mass $m^{\ast }$ will depend on $M$.
This is because the polaron mass $m^{\ast }=m\exp (A/\omega )$ \cite{ale3},
where $m$ is the band mass in the absence of the electron-phonon
interaction, and $A$ is a constant. Hence, there is a large isotope effect
on the {\em carrier mass} in polaronic metals, in contrast to the zero
isotope effect in ordinary metals. Recently, this effect has been
experimentally found in cuprates \cite{zao} and manganites \cite{pet}. I
anticipate the same effect in MgB$_{2}$. Also the optical conductivity of
polarons is different from ordinary electrons. In particular, there is a
substantial non-Drude polaron midinfrared conductivity at the expense of the
Drude contribution. On the other hand the nonadiabatic polaron $dc$
conductivity has a metallic character if the temperature is below the
characteristic optical phonon frequency. Its magnitude and temperature
dependence are determined by  the coupling with low-frequency phonons alone (like in
usual metals) because there is no scattering off high-frequency phonons
bound with the carriers into coherent (Glauber) polaronic states.

The auhtor is  grateful to Alex Bratkovsky and Viktor Kabanov for
illuminating discussions of the band structure and physical properties
of cuprates and MgB$_2$.

\begin{table}[tbp]
\caption{The Fermi energy (multiplied by the degeneracy) of cuprates
  and MgB$_2$}
\begin{tabular}[t]{llllllll}
Compound & T$_{c}$ (K) & $\lambda _{H,ab}\cite{alekabT}$ $(\AA )$ & 
d$(\AA )$ & $gE_{F}$ (meV) &  &  &  \\ \hline
$La_{1.8}Sr_{0.2}CuO_{4}$ & 36.2 & 2000 & 6.6 & 112 &  &  &  \\ 
$La_{1.78}Sr_{0.22}CuO_{4}$ & 27.5 & 1980 & 6.6 & 114 &  &  &  \\ 
$La_{1.76}Sr_{0.24}CuO_{4}$ & 20.0 & 2050 & 6.6 & 106 &  &  &  \\ 
$La_{1.85}Sr_{0.15}CuO_{4}$ & 37.0 & 2400 & 6.6 & 77 &  &  &  \\ 
$La_{1.9}Sr_{0.1}CuO_{4}$ & 30.0 & 3200 & 6.6 & 44 &  &  &  \\ 
$La_{1.75}Sr_{0.25}CuO_{4}$ & 24.0 & 2800 & 6.6 & 57 &  &  &  \\ 
$YBa_{2}Cu_{3}O_{7}$ & 92.5 & 1400 & 4.29 & 148 &  &  &  \\ 
$YBaCuO(2\%Zn)$ & 68.2 & 2600 & 4.29 & 43 &  &  &  \\ 
$YBaCuO(3\%Zn)$ & 55.0 & 3000 & 4.29 & 32 &  &  &  \\ 
$YBaCuO(5\%Zn)$ & 46.4 & 3700 & 4.29 & 21 &  &  &  \\ 
$YBa_{2}Cu_{3}O_{6.7}$ & 66.0 & 2100 & 4.29 & 66 &  &  &  \\ 
$YBa_{2}Cu_{3}O_{6.57}$ & 56.0 & 2900 & 4.29 & 34 &  &  &  \\ 
$YBa_{2}Cu_{3}O_{6.92}$ & 91.5 & 1861 & 4.29 & 84 &  &  &  \\ 
$YBa_{2}Cu_{3}O_{6.88}$ & 87.9 & 1864 & 4.29 & 84 &  &  &  \\ 
$YBa_{2}Cu_{3}O_{6.84}$ & 83.7 & 1771 & 4.29 & 92 &  &  &  \\ 
$YBa_{2}Cu_{3}O_{6.79}$ & 73.4 & 2156 & 4.29 & 62 &  &  &  \\ 
$YBa_{2}Cu_{3}O_{6.77}$ & 67.9 & 2150 & 4.29 & 63 &  &  &  \\ 
$YBa_{2}Cu_{3}O_{6.74}$ & 63.8 & 2022 & 4.29 & 71 &  &  &  \\ 
$YBa_{2}Cu_{3}O_{6.7}$ & 60.0 & 2096 & 4.29 & 66 &  &  &  \\ 
$YBa_{2}Cu_{3}O_{6.65}$ & 58.0 & 2035 & 4.29 & 70 &  &  &  \\ 
$YBa_{2}Cu_{3}O_{6.6}$ & 56.0 & 2285 & 4.29 & 56 &  &  &  \\ 
$HgBa_{2}CuO_{4.049}$ & 70.0 & 2160 & 9.5 & 138 &  &  &  \\ 
$HgBa_{2}CuO_{4.055}$ & 78.2 & 1610 & 9.5 & 248 &  &  &  \\ 
$HgBa_{2}CuO_{4.055}$ & 78.5 & 2000 & 9.5 & 161 &  &  &  \\ 
$HgBa_{2}CuO_{4.066}$ & 88.5 & 1530 & 9.5 & 274 &  &  &  \\ 
$HgBa_{2}CuO_{4.096}$ & 95.6 & 1450 & 9.5 & 305 &  &  &  \\ 
$HgBa_{2}CuO_{4.097}$ & 95.3 & 1650 & 9.5 & 236 &  &  &  \\ 
$HgBa_{2}CuO_{4.1}$ & 94.1 & 1580 & 9.5 & 257 &  &  &  \\ 
$HgBa_{2}CuO_{4.101}$ & 93.4 & 1560 & 9.5 & 264 &  &  &  \\ 
$HgBa_{2}CuO_{4.101}$ & 92.5 & 1390 & 9.5 & 332 &  &  &  \\ 
$HgBa_{2}CuO_{4.105}$ & 90.9 & 1560 & 9.5 & 264 &  &  &  \\ 
$HgBa_{2}CuO_{4.108}$ & 89.1 & 1770 & 9.5 & 205 &  &  &  \\ 
$MgB_{2}$ & 39 & 1400 \cite{lambda} & 3.52& 122 &  &  &  \\ \hline
\end{tabular}
\end{table}

\end{document}